# Calculation of Evaluation Variables for High Gradient Magnetic Separation with an Idealized Capture Model

Fengyu Xu, Anbin Chen
*School of Electrical Engineering and Automation, Harbin Institute of Technology, Harbin, 150006, China*

*Abstract*—This paper regards feed mine as a mixture of intergrowths and pure non-magnetic mineral particles, presents a method to calculate the evaluation variables such as grade and recovery in high gradient magnetic separation (HGMS). A idealized capture model is constructed in which the interaction between particles is not taken into account and only for the initial aggregation condition that the separator has the highest capture efficiency. In the model we adopt the functions that use nominal particle radius and magnetic mineral content as independent variables to describe volume fraction distribution and capture efficiency of intergrowths respectively. Through adding multi-wire magnetic fields and setting periodic boundary conditions in flow field analysis, we modify the computational domain of the single-wire capture theory to a element domain that periodically appears in the multi-wire matrix. By means of finite element software, particle trajectories, flow field and magnetic field are clearly exhibited, and then capture efficiency function is obtained by interpolation method. The calculated evaluation variables theoretically represent the best performance of magnetic separator for a given feed. They can assist mineral engineers to evaluate or compare the effects of different magnetic separation systems in advance. We use removal of iron impurity from kaolin as an example to illustrate the presented calculation method. The results quantitatively compare the evaluation variables of the separation at different magnetic fields and show that the advantage of higher magnetic field in separation efficiency decreases with the increase of saturation magnetization of magnetic mineral.

*Index Terms*—High gradient magnetic separation, mineral processing, particle capture, intergrowths, magnetic field, laminar flow.
PACS—41.20.Gz, 47.57.J-, 47.85.M-.

## I. Introduction

THERE are some known interactions between mineral particles, such as agglomeration, collision, competitive accumulation, and entrainment, which have negative effects on the separation capability of high gradient magnetic separators. The existence of these effects has been confirmed in theory and in practice[1-8]. So far, however, it is almost entirely dependents on mineral processing experiments, in order to determine whether a specified separator configuration has economic value for separating a particular mineral, and the supports from the magnetic separation theory are often limited qualitative analysis [6-15]. Though the precise mathematical descriptions are already available for all the interactions between the mineral particles, but it is hard to bear for solving an enormous number of equations that describe the interaction forces between particles in a meaningful numerical simulation. Therefore, mineral engineers have to organize a large number of experiments to attain a valuable separator's performance, although it is often vague where the best "point" should be. It is well known that, all interactions play negative roles in the frame of HGMS theory. If don't take these interactions into account, we can at least answer such question in theory that for a given mixture of mineral particles, what is the best separation performance the separator can achieve? This is the motivation to propose an idealized capture model in the paper.

In multi-wire magnetic filter theory, the phenomenon based models that were developed by J.H. P. Watson et al. [16-18] has assumed that the particles are composed of the single magnetic mineral and are of a single particle size. Therefore, the theory can be satisfactorily applied to the occasions where particles are often made artificially and particle size distribution is relatively concentrated [9-10,19]. In these occasions, the nature of the magnetic separation mostly is to separate magnetic particles from slurry solution. But in the mineral processing occasions, the nature is to separate the magnetic particles from the non-magnetic particles, or to separate the relatively strong magnetic particles from the relatively weak magnetic particles. The compositions of the particles are much more complex than the artificial particles. For these particles, not only the particle size is various, but also the degree of liberation is. The mineral particle that is not completely liberated, in other words, that includes magnetic mineral and non-magnetic mineral simultaneously, are known as intergrowth. It has been proved that the composition and distribution of intergrowths are the important factors that determine upper limit of separator's performance. X. Zheng etc. [7] have studied the behavior of intergrowths and illustrated the existence of competitive accumulation between them. In our particle capture model, the goal is to calculate evaluation variables of the separator such as grade and recovery, the behaviors of all intergrowths must be fully considered.

The concept of capture radius is the basis for calculating HGMS separation efficiency, and its numerical value is determined by solutions of particle motion equation, in which force fields must be firstly solved. Instead of using analytical formula, some researchers tend to use numerical methods to calculate the force fields [23-26]. This can at least lead to a

e-mail: xufy@hit.edu.cn

benefit that allows to introduce the effects of multi-wire on the magnetic field and flow field into the calculation model and that makes capture radius more representative in a real high gradient magnetic separator. Specially, with the progress of particle tracing technology in the commercial finite element (FE) software, the numerical calculation of the magnetic particle trajectory is becoming more and more convenient. When the particle size is small enough, the reaction of particles on the magnetic field and flow field can be neglected. Under the assumption of ideal capture model, we calculate the clusters of motion trajectories for the particles with discrete sizes and discrete magnetic properties by using COMSOL software, then determine the capture radii and the capture efficiency matrix in turn, and then convert the matrix into the continuous function by using two-dimensional interpolation method, finally the calculation formulae of recovery and grade are derived. Because all sorts of magnetic particles are treated equally in the model, the calculated results reflect competitive accumulation.

According to the single-wire capture theory, as the buildup of particles on the wire increases, the capture radius will decrease, until the capture radius corresponds to the buildup surface [18]. It can be inferred that the ferromagnetic wire has the best capture ability in the initial aggregation stage. At this time the influence of the buildup on the flow field can be neglected, and the calculation model is simplified.

We use kaolin clay as an example to illustrate the presented calculation method. In fact, the magnetic mineral content of kaolin is very low, and the solid concentration of slurry is low during magnetic separation as well. Therefore there are fewer opportunities for interaction between mineral particles, which is close to the idealized capture model. In the example we take saturation magnetization with zero and the maximum measurement value respectively for the magnetic mineral in kaolin - hematite, and set the magnetic field of the HGMS machine to 1 Tesla and 5 Tesla respectively, therefore there are four different cases in total. Through the calculation for these cases, the paper attempts to quantitatively explain a complaint of some kaolin mine engineers from the theoretical point of view. That is, in the situation of same pre-treatment kaolin slurry, the magnetic separation efficiency of 5 Tesla separator by one pass is only similar to that of 1 Tesla separator by two passes.

## II. THE TWO-VARIABLES VOLUME FRACTION DISTRIBUTION OF INTERGROWTHS

The grade and the recovery are primary variables to evaluate the effectiveness of magnetic separator for a given feed material. The experiences show that they not only depend on the parameters of magnetic separator, but also depend on the properties of mineral particles, such as particle size distribution, magnetic properties and degree of dissociation [20-22]. According to the particle capture theory, when the magnetic force and fluid drag force are dominant, the motion behavior of magnetic particles only depend on its particle size and magnetic mineral content for a given HGMS's configuration. Representing the nominal particle radius by $b$, and the magnetic mineral content in the particles by $\zeta$, this paper regards the volume fraction distribution of magnetic particles as a function of these two variables, and divide all feed particles into two groups, one is only containing non-magnetic minerals corresponding to $\zeta=0$, the other is the intergrowths with different magnetic mineral content corresponding to $\zeta \in (0, 1]$. Obviously, the monomers of the magnetic mineral are contained in the intergrowths here. Because the original definition of magnetization is the total magnetic moment per unit volume, in order to make the expression of the magnetic field force more intuitive, all the mass fractions are converted into volume fraction in the calculation model. The volume fraction distribution of intergrowths in feed can be decomposed into three parts as follows.

### A. The Particle Size Distribution in Feed

Regarding all particles in the feed as the overall sample in the view of probability theory, let $p_{\text{fd}}(b)$ denote the feed particle size distribution in volume fraction, which can be measured with a particle size analyzer. In general, according to different characteristics of particle size distribution, normal, lognormal, and Rosin-Rammler distributions can be used for approximate analytical functions. In order for higher accurate, we use a spline interpolation method to construct the continuous form of the distribution function. Whichever form of function is used, in the particle size range of $b \in [b_{\min}, b_{\max}]$, the function $p_{\text{fd}}(b)$ satisfies

$$\int_{b_{\min}}^{b_{\max}} p_{\text{fd}}(b)\, db = 1. \qquad (1)$$

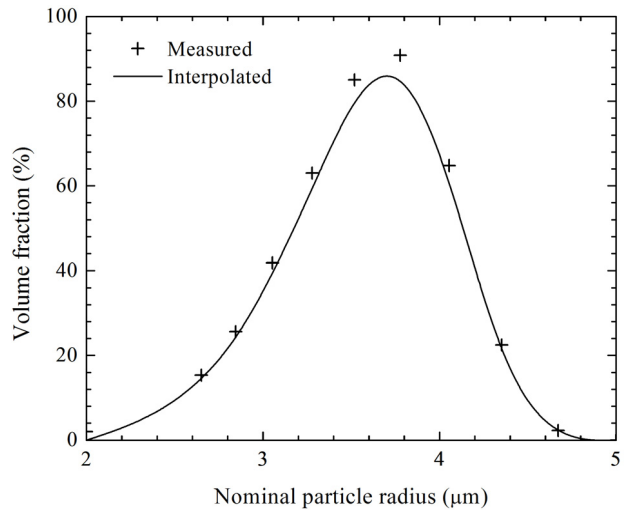

Fig. 1 The interpolation function of the particle size distribution for a kaolin feed

We take a group of kaolin data measured by particle size analyzer. The data is transformed from mass fraction to volume fraction and from discrete to continuous in turn, and then used for linear spline interpolation. Fig. 1 shows a plot of the particle size distribution function. In the figure the plus sign indicates the measured data point after transformations, the solid line is the curve of the interpolation function, and the range of the nominal particle radius is $b \in [2\mu m, 5\mu m]$.

### B. Magnetic Mineral Content Distribution of Intergrowths with an Arbitrary Nominal Particle Radius

Regarding all intergrowths with an arbitrary nominal particle radius as the overall sample in the view of probability theory, let $p_{mag}(b, \zeta)$ denote the magnetic mineral content distribution. Obviously, it is difficult to measure the raw data for this function. So far we have not found any corresponding instrument in the market. However, for the particles in which the fine grain magnetic minerals are embedded, the function can be derived from the mathematical point of view.

Firstly, we notice such fact that the shapes of the real mineral particles or grains are various, just are regarded as spherical in the capture model. So, in order to know how many magnetic grains can be embedded in a mineral particle at most, we just simply use their volume ratio. We assume that the nominal particle radii of the magnetic grains are approximately equal and denoted by $b_0$, then the maximum number of the embedded magnetic grains for the particles with nominal radius $b$ is

$$n = \text{floor}\left\{\left(\frac{b}{b_0}\right)^3\right\} \quad (2)$$

where floor$\{\cdot\}$ indicates rounding down. When the amount of the embedded fine magnetic grains is large, according to the Large Number Theorem in probability theory $p_{mag}(b, \zeta)$ follows a normal distribution

$$p_{mag}(b,\zeta) = \frac{C_{mag}}{\sqrt{2\pi}\sigma_\zeta} \exp\left\{\frac{(\zeta - \zeta_g)^2}{2\sigma_\zeta^2}\right\} \quad (3)$$

where $\zeta_g$ is mathematical expectation of $\zeta$, $\sigma_\zeta$ is variance, and $C_{mag}$ is defined to satisfy

$$\int_0^1 p_{mag}(b,\zeta)d\zeta = 1. \quad (4)$$

According to (2), when $b = \sqrt[3]{n}$, the intergrowths may contain from 1 to $n$ magnetic grains, the magnetic mineral content can be expressed as $\zeta \in [1/n, 1]$. Let us take $\zeta_g = \frac{1+1/n}{2}$ and $3\sigma_\zeta = \frac{1-1/n}{2}$, eqn. (3) becomes function $p_{mag}(n, \zeta)$. When different values of $n$ are taken, the curves of the distribution function are plotted in Fig. 2. The curves are consistent with the realistic scenario, where the magnetic grains content in the intergrowth tends to one as the particle size decreases. Substituting $n = (b/b_0)^3$ into $\zeta_g$ and $\sigma_\zeta$ we can obtain the continuous function $p_{mag}(b, \zeta)$. According to X. Yang's electron microscopy measurements [27], we take the nominal radius of hematite grain equal to 0.4 μm. Corresponding to the range of nominal radius $b \in [2\mu m, 5\mu m]$, the minimum embedded grain number is $n_{min}=125$, the maximum number is $n_{max}=1953$. It can be seen from Fig. 2 that, when $n$ is large enough, $p_{mag}(n, \zeta)$ approximately follows a normal distribution with $\zeta_g=1/2$ and $\sigma_\zeta=1/6$.

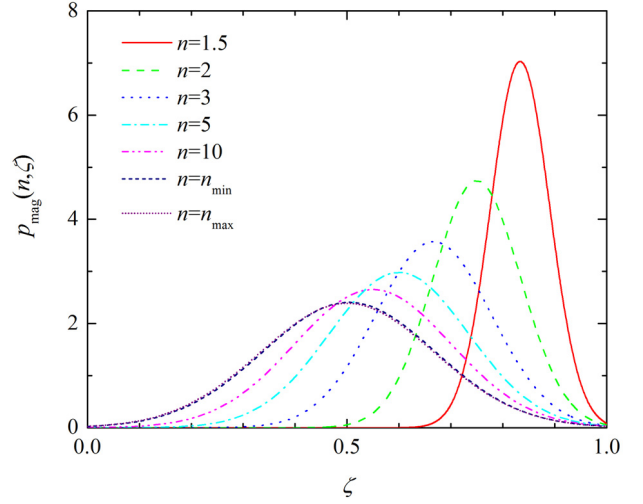

Fig. 2 The function $p_{mag}(n, \zeta)$ when different value of $n$ is taken. $n_{min}=125$, $n_{max}=1953$.

### C. Particle Size Distribution of Intergrowths in Feed

Regarding all particles in the feed as the overall sample of probability theory, let $p_{itg}(b)$ denote particle size distribution of intergrowths, which is hard to measure directly because it need to separate all the intergrowths from the feed before measuring. We still consider the fine magnetic grains embedded intergrowth, and assume that the opportunity for each grain to embed arbitrary intergrowth is equal. The particle size distribution of intergrowths in feed should follow the average distribution, which means that the function $p_{itg}(b)$ is a constant independent of particle size. Let $p_{itg}(b)=C_{itg}$, it can be calculated by mass balance equation of magnetic minerals in feed

$$\int_{b_{min}}^{b_{max}} p_{fd}(b) C_{itg} \int_0^1 \zeta p_{mag}(b,\zeta) d\zeta db = \alpha_v, \quad (5)$$

where $\alpha_v$ is volume fraction of total magnetic mineral in feed, which can be measured by chemical method but the measurement result is usually represented by mass fraction, the transform relationship between them is

$$\alpha_v = \frac{\alpha_m/\rho_{mag}}{\alpha_m/\rho_{mag} + (1-\alpha_m)/\rho_{gan}}, \quad (6)$$

where $\alpha_m$ is mass fraction of total magnetic mineral in feed, $\rho_{mag}$ is density of magnetic mineral, $\rho_{gan}$ is density of non-magnetic mineral.

Based on the above analyses, the two-dimensional volume fraction distribution of intergrowths in feed can be expressed as

$$p_{mag/fd}(b,\zeta) = p_{fd}(b) \cdot C_{itg} \cdot p_{mag}(b,\zeta). \quad (7)$$

In our example, we take $\alpha_m=1.0\%$, along with the density of the hematite $\rho_{mag}=5.26$ g/cm$^3$ and the density of the kaolin $\rho_{gan}=2.6$ g/cm$^3$, then $C_{itg} = 9.936\times10^{-3}$ is solved from (5). Thus the continuous distribution function of $p_{mag/fd}(b, \zeta)$ is shown in Fig. 3. It clearly shows that iron impurities are mainly distributed around the point with $b$=3.5 μm, $\zeta$=0.5.

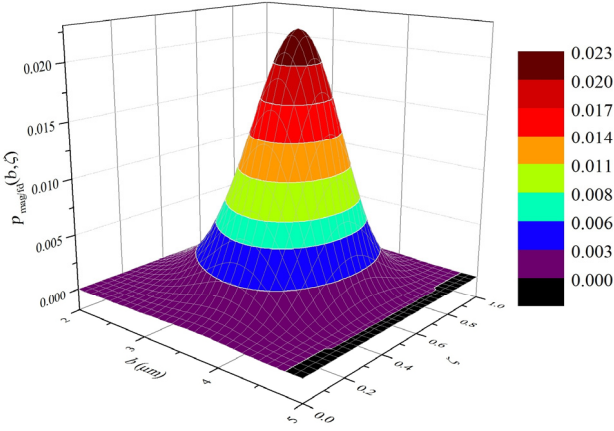

Fig. 3 The two-dimensional volume fraction distribution of intergrowths in a kaolin feed. $b$ - nominal particle radius, $\zeta$ - magnetic mineral content in an intergrowth.

## III. THE IDEALIZED PARTICLE CAPTURE MODEL

### A. Computational Domains

We use COMSOL software to calculate the capture efficiency matrix for all kinds of intergrowths in a given feed. The first issue is to determine the computational domains. The single-wire particle capture theory points out that the particle trajectory will show an oscillating behavior for a high value of Stokes number. In order to the capture efficiency calculation is not influenced by the oscillating behavior, the computational domains and the coordinate system in the wire's cross-sectional plane are arranged as Fig. 4. The Stokes number should be checked by the following formula [18]

$$S = 2\rho_p b^2 |V_0|/(9\eta a), \quad (8)$$

where $V_0$ is fluid velocity, $\eta$ is slurry viscosity, $a$ is radius of ferromagnetic wire, $\rho_p$ is density of particle, which can be expressed as

$$\rho_p = \frac{\zeta \rho_{mag}}{\zeta \rho_{mag} + (1-\zeta)\rho_{gan}} \quad (9)$$

It is easy to know that $\rho_P \in [\rho_{gan}, \rho_{mag}]$. In our example, we take the wire diameter $2a$=50 μm and the fluid velocity $V_0$=10 mm/s on the inlet boundary, the calculated range of the Stokes number is $S \in [5.2 \times 10^{-4}, 6.6 \times 10^{-3}]$, which is far smaller than the boundary between high value and low value S=1/8 [18]. Therefore the target computational domain of particle trajectories is set to Zone III, but in order to avoid particles interact with parallel boundaries, the real computational domain is set to Zone II-IV. Because the computational domain of flow field must cover that of the particle trajectory and the influence of the adjacent wires on the flow field requires setting periodic boundary conditions on the horizontal walls, therefore the computational domain of the flow field is set to Zone I-IV. In principle, it is enough for the computational domain of magnetic field to cover the target computational domain of particle trajectories, so the computational domain of the magnetic field is set to Zone III. In order to make the magnetic field more generally represent the situation in the multi-wire matrix, in addition to the wires within the Zones, the contributions of the adjacent wires on the magnetic field in Zone III we also take into account. All involved wires as shown in Fig. 5. The magnetic filed in Zone III is a vector superposition of the magnetic fields that are generated by these wires, which will be described in detail below. The gravitational field is ignored because it is very small compared to the fluid drag force.

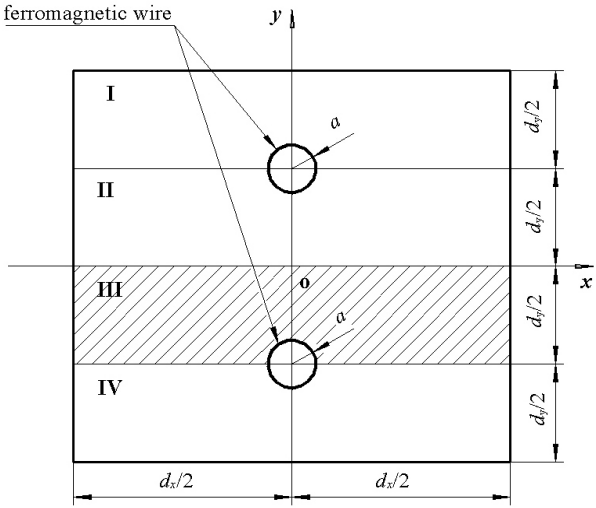

Fig. 4 The computational domains of particle trajectories.

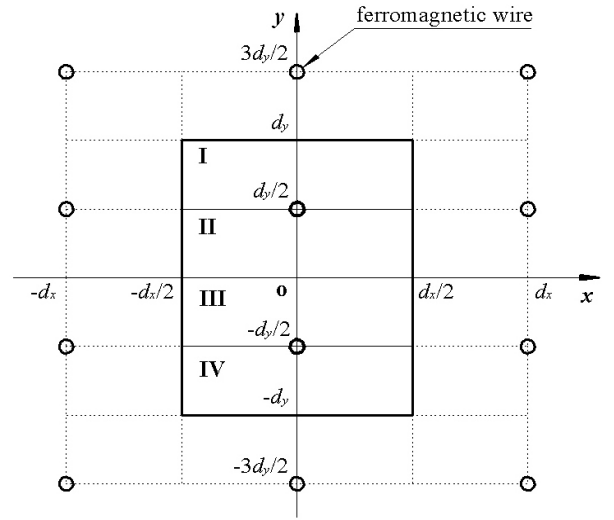

Fig. 5 The ferromagnetic wires used to calculate the magnetic field in Zone III.

## B. Flow Field Calculation

The Navier-Stokes laminar flow equations are used for conservation of momentum and the continuity equation for conservation of mass. In the model, the left boundaries of the computational domains are set to inlet, the right boundaries are set to outlet, the inflow velocity on the inlet is set to $V_0$=10 mm/s, and the pressure on the outlet is set to 0. Both horizontal boundaries are set as periodic boundary conditions. In Fig. 4 and Fig. 5 we take the distance between vertically adjacent wires $d_y$=10$a$, and the distance between horizontally adjacent wire $d_x$=20$a$. They should satisfy the condition that the disturbance of the fluid caused by upstream wire almost disappear before the fluid enters into the acting zone of the downstream adjacent wire.

Specially, slurry viscosity is a parameter that should be appropriately set. It is well known that the slurry viscosity increases with the increase of concentration, and when the concentration is equal to 0 it is equal to the water viscosity $1\times10^{-3}$ Pa·s. For kaolin slurry, combined with the two data points published by [28], which are $\eta$=4×10$^{-3}$ Pa·s at the concentration of 30% and $\eta$=20×10$^{-3}$ Pa·s at the concentration of 70%, we use linear spline interpolation function to approximate the relationship between the viscosity and the concentration as shown in Fig. 6. In general, the solid concentration of the kaolin slurry is 15%, so we get the slurry viscosity $1.777\times10^{-3}$ Pa·s from the figure. By neglecting the influence of impurities, the density of the kaolin slurry is expressed as

$$\rho_{slurry} = C \times \rho_{gan} + (1-C) \times \rho_f \quad (10)$$

where $C$ is the slurry concentration, $\rho_f$ is the density of water, and $\rho_f$=1.0 gm/cm$^3$, then the calculated value of the slurry density is $\rho_{slurry}$=1.24 g/cm$^3$.

The calculated result of the flow field is shown in Fig. 7, where the color represents amplitude of local flow velocity, the black solid line represents streamline. It can be seen from the streamline, the disturbance caused by the wire is almost disappear before flow arrives at the outlet boundary.

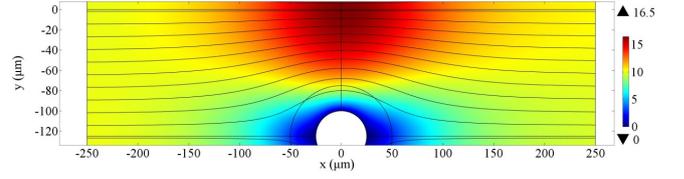

Fig.7 The flow field in Zone III, where the inflow velocity $V_0$=10 mm/s.

In the figure, the maximum amplitude of flow velocity $V_{max}$=16.5mm/s. According to Reynolds Number formula[16]

$$Re = (\rho_{slurry} V_{max} \cdot 2a)/\eta \quad (11)$$

the maximum Reynolds Number $Re$ = 0.576 < 1, satisfies the condition to calculate laminar flow field.

## C. Magnetic Force Field Calculation

Due to contain a small amount of ferromagnetic minerals, hematite grains often show magnetic saturation. The known measurement data show that the saturation magnetization of hematite increases with the decreases of the particle size[17, 29]. X. Yang has proposed that the nominal diameter is in the range of 0.3~0.8 μm for the hematite grains that are embedded in an intergrowth of kaolin clay [27]. For the purpose of simplification, we assume the grain sizes of the embedded hematite are equal, then the magnetization is proportional to the hematite content in an intergrowth. Therefore, the magnetization of an intergrowth can be expressed as

$$M_{itg} = \zeta M_{FeO} = \zeta(M_{F0} + \kappa_{FeO} H) \quad (12)$$

where $M_{FeO}$ is magnetization of hematite grains, $M_{F0}$ is saturation magnetization of hematite grains, and $\kappa_{FeO}$ is the volume magnetic susceptibility after the saturation point. The radial and tangential components of the magnetic force that acts on the intergrowth around ferromagnetic wire respectively are

$$F_{mr} = \frac{1}{2}\mu_0 V_p \zeta \left[ M_{F0}\nabla_r H + \kappa_{FeO}\nabla_r(H^2) \right] \quad (13a)$$

$$F_{m\theta} = \frac{1}{2}\mu_0 V_p \zeta \left[ M_{F0}\nabla_\theta H + \kappa_{FeO}\nabla_\theta(H^2) \right] \quad (13b)$$

where $\mu_0$ is vacuum magnetic susceptibility, $V_p$ is the particle volume, and $V_p = \frac{4}{3}\pi b^3$ according to the assumption of the spherical shape. If consider $V_p$ and $\zeta$ as a whole, (13) implies a fact that only magnetic mineral content contributes to the magnetic force that acts on the particle in an intergrowth.

If the magnetic strength $H$ in (13) is calculated by finite element software, the gradients of H and its square can be calculated by finite difference method. The advantage of this method is to allow to calculate for the wires with various cross-sectional shapes, for example, circle, rectangle, diamond, etc. The disadvantage is to tend to introduce a large computational error in the regions where magnetic field varies sharply. In order to compare with existing single wire capture theory, we choose the wire with circular cross section to calculate magnetic force. Only considering the situation of saturation magnetized wire, we use the following analytic formulae to calculate two components of the magnetic force [7]

$$F_{mr} = -2\mu_0 V_p A \frac{a^2}{r^3} \zeta [fM_{F0} + \kappa_{FeO} H_0] H_0 \left( \frac{Aa^2}{r^2} + \cos 2\theta \right)$$
(14a)

$$F_{m\theta} = -2\mu_0 V_p A \frac{a^2}{r^3} \zeta [fM_{F0} + \kappa_{FeO} H_0] H_0 \sin 2\theta \quad (14b)$$

where

$$A = M_{wr}(H_0)/(2H_0),$$
$$f = \frac{1}{2\sqrt{1 + A^2 a^4/r^4 + 2A(a^2/r^2)\cos 2\theta}},$$

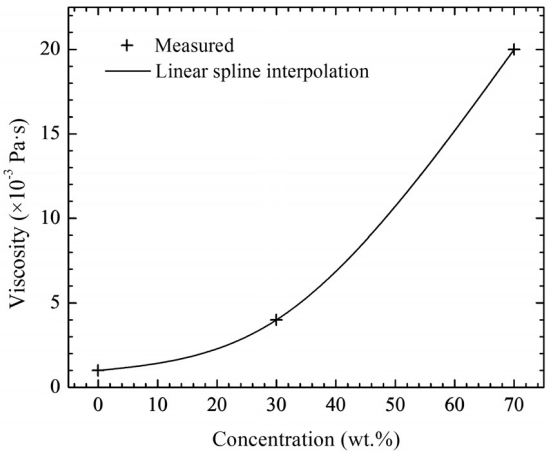

Fig. 6 The interpolation function of viscosity versus concentration for kaolin slurry.

$M_{wr}(\cdot)$ is magnetization function of ferromagnetic wire which is instead of a constant number to precisely calculate, $H_0$ is applied magnetic field strength. For the wire made of SUS430, the magnetization function can be transformed from a measured B-H curve [30]. Compared with the finite difference method, the advantage of the analytical formulae is easy to obtain smooth and accurate magnetic field distribution. When $M_{F0}=0$, eqn. (14) is reduced to [18]

$$F_{mr} = -\mu_0 V_p M_{wr}(H_0) \zeta \kappa_{FeO} \frac{a^2}{r^3} \left( \frac{M_{wr}(H_0)a^2}{2r^2} + H_0 \cos 2\theta \right)$$
(15a)

$$F_{mr} = -\mu_0 V_p M_{wr}(H_0) \zeta \kappa_{FeO} \frac{a^2}{r^3} H_0 \sin 2\theta \quad (15b)$$

In the software, we perform the following steps to calculate the magnetic forces for an intergrowth represented by number pairs $(b, \zeta)$. At an arbitrary spatial point with coordinate $(x, y)$,

Step 1 take circle center of each wire in the cross sectional plane as origin of Polar coordinate system, apply eqn. (14) to calculate the radial and circumferential components of magnetic force.

Step 2 sum the components of magnetic force on radial and circumferential directions respectively in the coordinate system with the origin of $O$ as shown in Fig. 4.

Step 3 transform the component summation of magnetic forces from Polar coordinate system into Cartesian coordinate system $xoy$.

### D. Particle Trajectories Calculation

We divide the left boundary of Zone III into $N$ equal parts, let particles enter the zone at the midpoints of these parts with the velocity as same as the inflow velocity. The bottom midpoint is named as the 1st midpoint, and so on until the $N$th midpoint. In a solution $N$ particle trajectories can be calculated simultaneously.

According to the measurement data of L. Zheng, the saturation magnetizations of hematite $M_{F0}$ varies from about $1.05\times10^3$A/m to $1.394\times10^4$A/m; but the volume magnetic susceptibility $\kappa_{FeO}$ is approximately the same, equal to $1.852\times10^{-3}$ [29]. For an example, we consider a kaolin intergrowth with $b=3.5\mu m$ and $\zeta=0.5$. When $M_{F0}=0$ is taken, the calculated results of the particle trajectories, the magnetic field and the flow field are shown in Fig. 8, where $N=25$ is taken. Similarly, when $M_{F0}=1.394\times10^4$A/m is taken, the calculated results of the particle trajectories, the magnetic field and the flow field are shown in Fig. 9. In both figures, the red arrows indicate the directions of local magnetic forces that act on the particles, the blue solid lines indicate the contours of the magnetic forces, and the blue numbers represent locally the value of $\log(|F_M|/(\rho_p g))$, where $F_M$ is local magnetic force that acts on the particle, $g$ is gravitational acceleration, $|\cdot|$ indicates solving vector norm. In other words, the amplitude of local magnetic force is characterized by logarithm of ratio between vector norms of local magnetic force and local gravity force. By combining directions and contours of magnetic forces together, the figures can help us to understand the particle behaviors such as changing motion direction, acceleration, and deceleration.

In both figures, the color of particle trajectory indicates the value of $|v_p - v_f|/|v_f|$, where $v_p$ is local velocity of particle, $v_f$ is local velocity of slurry. The corresponding relationship between colors and numerical values is defined by the palette on the right side of the figures, the number above the palette represents the maximum value, and the below one represents the minimum. Because $|v_p - v_f|/|v_f| > 1$ occurs only in the region where is very close to the wire surface, so set the maximum color-corresponding value in the palette to 1 in order to observe acceleration and deceleration of particles in flow field more clearly. In addition, the particle radius is taken as the scale of 1:1 to compare the sizes of the particle and the wire intuitively.

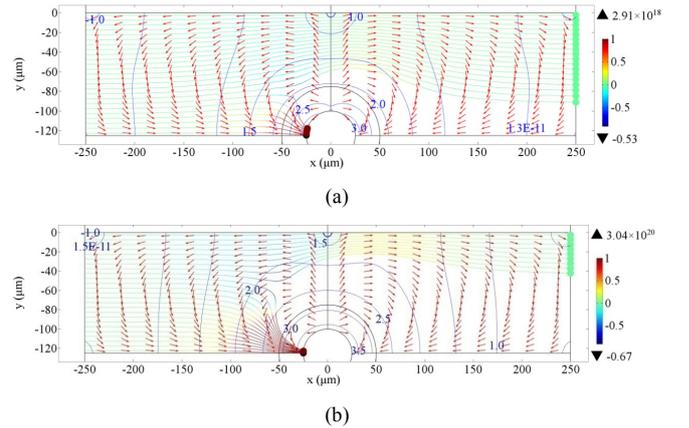

Fig. 8 The calculated results of the particle motion trajectories, the flow field and the magnetic field for a kaolin intergrowth with $M_{F0}=0$, $b=3.5\mu m$, $\zeta=0.5$ (a) when $B_0=1$T, (b) when $B_0=5$T.

We can find some facts from Fig. 8 and Fig.9 as follows. First, in the case of $M_{F0}=0$, when magnetic field increases from 1T to 5T, the number of captured particles changes from 7 to 16, increased by 128.6%; but in the case of $M_{F0}=1.394\times10^4$A/m, when magnetic field increases from 1T to 5T, the number of captured particles changes from 15 to 20, only increased by 33.3%. This tell us that the separation efficiency can be effectively improved by increasing the magnetic field from 1T to 5T only when the saturation magnetization of hematite is lower. Second, during the movement the particles that are not captured are repulsed by the magnetic force and move away from the wire. The greater the saturation magnetization of the hematite and the applied magnetic field, the more obvious this movement phenomenon is. One may readily see that, the particles does not follow the average probability to enter the capture zone of the downstream wire. If the wires in magnetic separator is aligned in the vertical direction as shown in Fig. 4, the particles that are not captured by upstream wire can hardly be captured by downstream adjacent wire. Therefore, in the initial stage of aggregation, the capture probability of the first wire column to the particles is that of the whole magnetic separator. On the contrary, if the wires of adjacent columns staggered in the vertical direction, the particles that are not captured by the upstream wire can be more easily captured by the downstream adjacent wire. Obviously, compared with the aligned arrangement, the staggered arrangement has a significant advantage in the separation efficiency. But for a fine ferromagnetic wire, e.g. wire diameter of 50μm presented in this paper, it is not easy to fabricate a wire stack with staggered form. Third, due to the spatial magnetic field force, the amplitude and direction of particle velocity constantly changes, which are even very severely in the region close to the wire. Therefore, there is an inevitable collision between intergrowths and pure non-magnetic mineral particles. These collisions will disturb the normal magnetic separation process, moreover deteriorate the capability of magnetic separator. In addition, if the distances between intergrowths are too short, magnetic agglomeration may occur. Since these interactions are not considered, we say that the proposed model is idealized and the calculated results can only represent the best separation results of magnetic separator. In fact, the engineers of mineral separation plants always attempt to make a compromise between separation efficiency and productivity by adjusting slurry concentration.

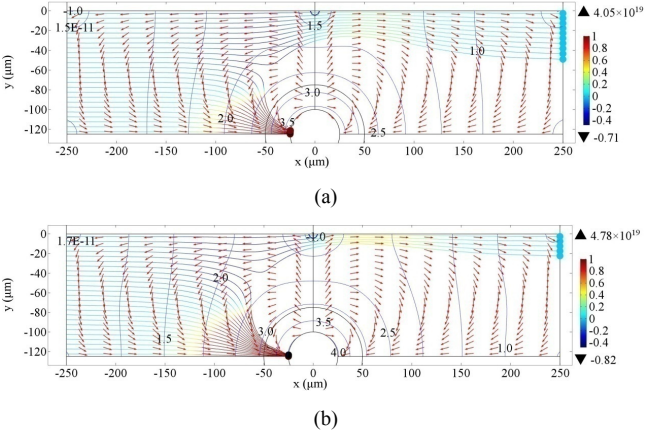

Fig. 9 The calculated results of the particle motion trajectories, the flow field and the magnetic field for a kaolin intergrowth with $M_{F0}=1.394\times10^4$A/m, $b=3.5$μm, $\zeta=0.5$ (a) when $B_0=1$T, (b) when $B_0=5$T.

Ideally, we assume that the particles obey the average probability distribution at the inlet boundary. Then, in the model of Fig. 4 the capture efficiency of intergrowths is equal to the ratio of the particles that eventually fell onto the wire to total particles that start from the inlet boundary. For all the intergrowths in the feed, the capture efficiency is a function of nominal particle diameter and magnetic mineral content, can be denoted as $E_{cap}(b, \zeta)$. In the Zone III shown in Fig. 4, it is defined as due to single wire capture theory

$$E_{cap}(b,\zeta) = R_c(b,\zeta)/(d_y/2) \qquad (16)$$

where $R_c$ represents capture radius. As can be seen from Fig. 8

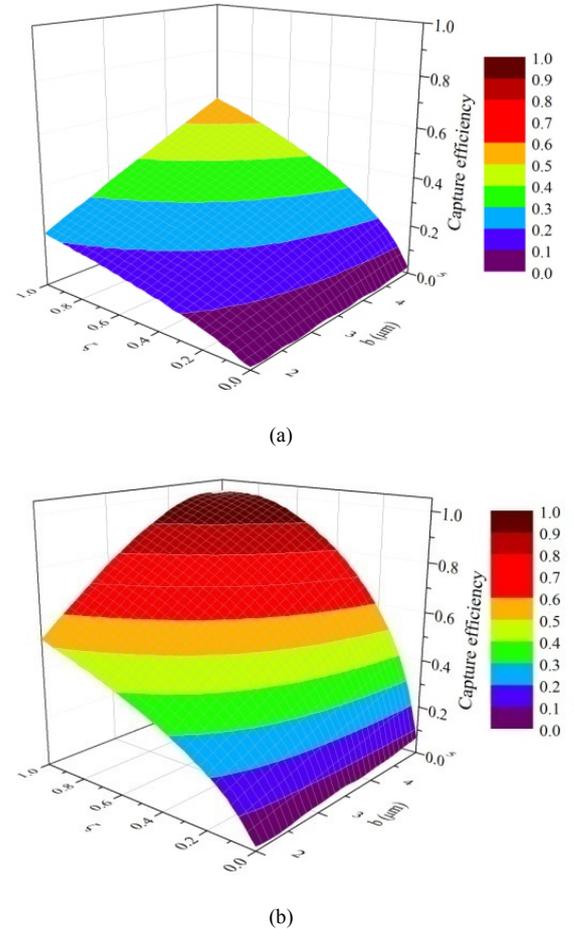

Fig. 10 The capture efficiency function of the kaolin intergrowths for SUS430 wire with diameter $2a=50$μm ($M_{F0}=0$, $b\in[2$μm$,5$μm$]$, $\zeta\in(0,1]$), (a) $B_0=1$T, (b) $B_0=5$T.

and Fig. 9, the greater the $N$ value is, the more accurate the calculated capture efficiency is, so we take $N=200$ in the below calculation. It is worth noting that the definition of the capture efficiency is only suitable for the initial stage of aggregation, which remarkable characteristic is that the aggregation of particles on ferromagnetic wire does not affect the trajectories of the particles.

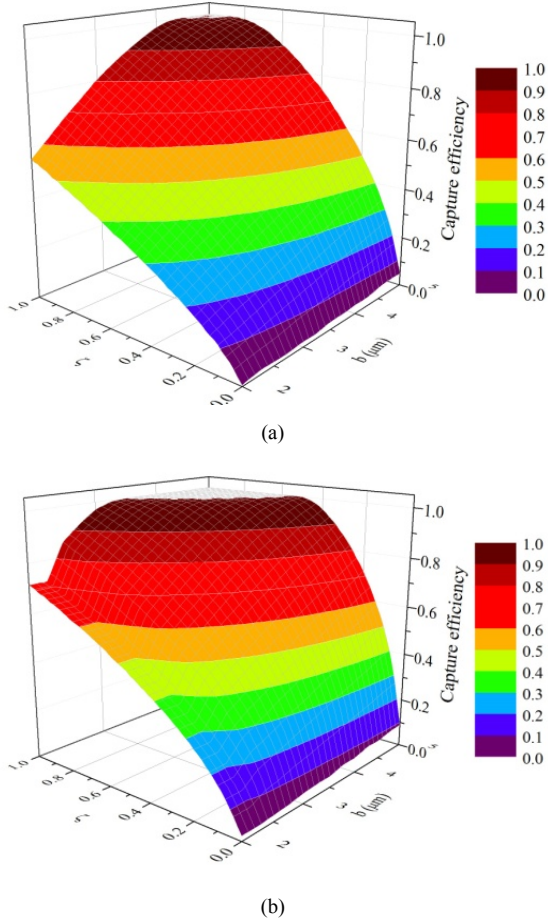

Fig. 11 The capture efficiency function of the kaolin intergrowths for SUS430 wire with diameter $2a$=50μm ($M_{F0}$=1.394×10$^4$A/m, $b\in$[2μm,5μm], $\zeta\in$(0,1]), (a) $B_0$=1T, (b) $B_0$=5T.

The discrete values of capture efficiency are calculated on the discrete data pairs ($b$, $\zeta$) using finite element software and the continuous function of capture efficiency is formulated with two dimensional linear spline interpolation. As an example, the capture efficiency function of the kaolin intergrowths in the feed are shown in Figure 10 and Figure 11, where the ferromagnetic wire is made of SUS430 with diameter $2a$=50μm, in Fig. 10 the saturation magnetization of hematite $M_{F0}$=0, and $M_{F0}$=1.394×10$^4$A/m in Fig. 11. For the purpose of comparison, each figure illustrate two cases where The background magnetic fields of $B_0$=1T and $B_0$=5T are applied respectively. These two figures show that the capture efficiency is a monotonically increasing function of particle radius and magnetic mineral content, and a higher $M_{F0}$ makes the capture efficiency function of the 1T magnetic field closer to that of the 5T magnetic field.

## IV. EVALUATION VARIABLES OF HGMS

### A. Evaluation Variables Calculation

Once the two-dimensional volume fraction distribution of intergrowths $p_{mag/fd}(b, \zeta)$ and the capture efficiency function of HGMS $E_{cap}(b, \zeta)$ are given, the grade and the recovery of separated minerals can be calculated. We consider the volume fraction of feed as 100%, the volume fraction of intergrowths in feed can be calculated by

$$V_{itg/fd} = \int_{b_{min}}^{b_{max}} \int_0^1 p_{mag/fd}(b,\zeta) d\zeta db. \tag{17}$$

The corresponding mass is

$$Q_{itg/fd}(0) = \int_{b_{min}}^{b_{max}} \int_0^1 [\zeta\rho_{mag} + (1-\zeta)\rho_{gan}] p_{mag/fd}(b,\zeta) d\zeta db, \tag{18}$$

where $Q_{itg/fd}(0)$ means the mass of intergrowths that are left in the feed after the 0th separation. The total mass of the pure non-magnetic particles is

$$Q_{gan} = (1 - V_{itg/fd})\rho_{gan}. \tag{19}$$

Since the pure non-magnetic particles can't be magnetically captured by ferromagnetic wire, no matter how many times the magnetic separations are, $Q_{gan}$ has remained the same. Thus, the total mass of the feed is

$$Q_{fd} = Q_{gan} + Q_{itg/fd}. \tag{20}$$

After $n$th separation, the mass of the captured intergrowths is

$$Q_{cpi}(n) = \int_{b_{min}}^{b_{max}} \int_0^1 [\zeta\rho_{mag} + (1-\zeta)\rho_{gan}] p_{mag/fd}(b,\zeta) \\ \cdot [1 - E_{cap}(b,\zeta)]^{n-1} E_{cap}(b,\zeta) d\zeta db. \tag{21}$$

the mass of the magnetic mineral in the captured intergrowths is

$$Q_{mag/cpi}(n) = \int_{b_{min}}^{b_{max}} \int_0^1 \zeta\rho_{mag} \cdot p_{mag/fd}(b,\zeta) \\ \cdot [1 - E_{cap}(b,\zeta)]^{n-1} E_{cap}(b,\zeta) d\zeta db. \tag{22}$$

Then the grade of total captured intergrowths is

$$\beta = \frac{\sum_1^n Q_{mag/cpi}(n)}{Q_{cpi}(n)} \tag{23}$$

and the recovery of the magnetic mineral is

$$\varepsilon = \frac{\sum_{1}^{n} Q_{mag/cpi}(n)}{Q_{fd} \cdot \alpha_{m}} \quad (24)$$

In the case of removing magnetic impurities from non-magnetic minerals, the concerned evaluation variables respectively are the mass fraction of residual magnetic mineral in non-magnetic mineral and the recovery of non-magnetic mineral. The non-magnetic mineral in the feed can be divided into two parts. One part involves all of pure non-magnetic particles, The other involves all of non-magnetic mineral contained in intergrowths. After $n$th separation, the mass of the non-magnetic mineral in the captured intergrowths is

$$Q_{gan/cpi}(n) = \int_{b_{min}}^{b_{max}} \int_{0}^{1} (1-\zeta)\rho_{gan} \cdot p_{mag/fd}(b,\zeta) \\ \cdot [1 - E_{cap}(b,\zeta)]^{n-1} E_{cap}(b,\zeta) d\zeta db. \quad (25)$$

The mass of the intergrowths that are left in the feed is

$$Q_{itg/fd}(n) = \int_{b_{min}}^{b_{max}} \int_{0}^{1} [\zeta\rho_{mag} + (1-\zeta)\rho_{gan}] p_{mag/fd}(b,\zeta) \\ \cdot [1 - E_{cap}(b,\zeta)]^{n} d\zeta db. \quad (26)$$

The mass of the magnetic mineral of intergrowths that are left in the feed

$$Q_{mag/lfi}(n) = \int_{b_{min}}^{b_{max}} \int_{0}^{1} \zeta\rho_{mag} \cdot p_{mag/fd}(b,\zeta) \\ \cdot [1 - E_{cap}(b,\zeta)]^{n} d\zeta db. \quad (27)$$

Therefore, the mass fraction of residual magnetic mineral is

$$\theta'(n) = \frac{Q_{mag/lfi}(n)}{Q_{gan} + Q_{itg/fd}(n)}. \quad (28)$$

The recovery of the non-magnetic mineral is

$$\varepsilon'(n) = 1 - \frac{\sum_{1}^{n} Q_{gan/cpi}(n)}{Q_{fd} \cdot (1-\alpha_{m})} \quad (29)$$

*B. The Calculated Results*

Removal of iron impurity from kaolin is used as our example, the distribution properties of the feed and the capture efficiency function are obtained in the 2nd and 3rd sections. The calculated evaluation variables of HGMS are listed in Table I and Table II, where the saturation magnetizations of hematite are $M_{F0}=0$ and $M_{F0}=1.394\times10^{4}$A/m respectively, the background magnetic intensity and the number of times of the feed passing through the separator are used as variables.

*C. Discussion*

As can be seen from the Table I and Table II, when the saturation magnetization of hematite $M_{F0}=0$, the separation efficiency at the 5 T magnetic field by 1 pass is better than that at the 1T magnetic field by 3 passes; when the saturation magnetization of hematite $M_{F0}=1.394\times10^{4}$A/m, the separation efficiency at the 5T magnetic field by 1 pass is similar to that at the 1T magnetic field by 2 passes. One may see that the hematite with higher saturation magnetization is easier to be removed, but makes the advantage of the 5T magnetic field degraded as well. Conversely, the 5T magnetic field has appropriate advantage than 1T only when the hematite has lower saturation magnetization, which seems to give a explanation for the complaint of those kaolin mine owners. We also note that in the magnetic separation practice of kaolin, it is hardly to obtain such a high value of recovery as high as that in Table I and Table II. It implies that, on the surface of ferromagnetic wire, the magnetic capture of magnetic particles is always accompanied by the mechanical capture of non-magnetic particles. In order to obtain more realistic evaluation variables, it is necessary to further study the mechanical capture.

The above analyses show that, with increase of the saturation magnetization of hematite the advantage of higher magnetic field decreases in separation efficiency, it can be explained by the magnetization properties of magnetic minerals in nature. From the original definition we can derive the following magnetic force formula for a hematite particle

$$F_{m} = \frac{1}{2}\mu_{0}V_{p}[M_{FeO}\nabla H - \kappa_{f}\nabla(H^{2})] \\ \approx \frac{1}{2}\mu_{0}V_{p}M_{FeO}\nabla H. \quad (30)$$

where $\kappa_{f}$ is the volume magnetic susceptibility of water.

TABLE I
THE CALCULATED EVALUATION VARIABLES, ASSUMING THE SATURATION MAGNETIZATION OF HEMATITE CONTENT IS EQUAL TO 0

| | $B_0$=1 T | | | $B_0$=5 T |
|---|---|---|---|---|
| | 1st pass | 2nd pass | 3rd pass | 1st pass |
| The hematite content in the feed (wt.%) | 1.0 | 1.0 | 1.0 | 1.0 |
| The mass fraction of residual magnetic mineral (%) | 0.71 | 0.51 | 0.268 | 0.33 |
| The recovery of non-magnetic mineral (%) | 99.9 | 99.8 | 99.7 | 99.7 |

TABLE II
THE CALCULATED EVALUATION VARIABLES, ASSUMING THE SATURATION MAGNETIZATION OF HEMATITE CONTENT IS EQUAL TO 1.394×10⁴A/M

| | $B_0$=1 T | | $B_0$=5 T |
|---|---|---|---|
| | 1st pass | 2nd pass | 1st pass |
| The hematite content in the feed (wt.%) | 1.0 | 1.0 | 1.0 |
| The mass fraction of residual magnetic mineral (%) | 0.35 | 0.14 | 0.17 |
| The recovery of non-magnetic mineral (%) | 99.7 | 99.6 | 99.6 |

It can be seen from (30), the magnetic force is proportional to the magnetization $M_{\text{FeO}}$. In the example of hematite, the magnetization can be expressed as a function of local magnetic field

$$M_{\text{FeO}}(B) = M_{F0} + (\kappa_{\text{FeO}}/\mu_0)B. \qquad (31)$$

When the background magnetic field increases from $B_1$ to $B_2$, the magnetization of hematite increases by

$$\frac{\Delta M_{\text{FeO}}}{M_{\text{FeO}}(B_1)} = \frac{B_2 - B_1}{\mu_0 M_{F0}/\kappa_{\text{FeO}} + B_1}. \qquad (32)$$

Let $B_1=1$T and $B_2=5$T, according to (32), the magnetization of hematite increases by 400% when $M_{F0}=0$; and the magnetization of hematite increases only by 38.2% when $M_{F0}=1.394\times10^4$ A/m. This is the essential reason why the advantage of higher magnetic field in separation efficiency decreases with the saturation magnetization increase of hematite.

## V. Conclusions

Based on the idealized capture model, we have presented a method to calculate the evaluation variables such as the grade and the recovery in high gradient magnetic separation (HGMS). The magnetic separation as the dominant physical process in HGMS has clarified. In the model, we have adopted functions of two independent variables, which are nominal particle radius and magnetic mineral content in a intergrowth, to describe volume fraction distribution of intergrowth and capture efficiency of ferromagnetic wire respectively. The intergrowth is referred to a particle characterized by nominal particle radii and magnetic mineral content in the paper. The calculated results represent theoretically the best performance of high gradient magnetic separator for a given feed mine. Thus, the objective of the single-wire capture theory is expanded from the particle group with same particle size and same kind of magnetic mineral to the mixture composed of intergrowths and pure non-magnetic particles. By means of finite element software, the model of the single-wire capture theory is modified to the element domain that is periodically appear in multi-wire matrix. On the basis of this research, the physical processes such as mechanical capture and collision could be continued to quantitatively study in breadth, and the change rule of evaluation variables during entire aggregation stage could be continued to quantitatively study in depth. These studies will enable us to more clearly recognize the complex physical processes of HGMS for a general mineral.

## Acknowledgment

The authors thank Professor Zhongyuan Sun who works at Central South University for useful discussions.

## References

[1] J.P. Glew and M.R. Parker. "The Influence of Interparticle Forces in the Magnetic Separation of Submicron Particles," *IEEE Trans. on Magnetics*, vol. Mag-20, no.5, 1984, pp. 1165-1167.
[2] J. Ren. "Dispersive Action of Sodium Silicate in Selective Agglomeration of Fine Hematite and Silicate Minerals Containing Iron," *Mining and Metallurgical Engineering* (in Chinese), vol. 11, no. 3, 1991, pp. 23-26.
[3] Q. Fang, J. Luo, J. Gan. "Interaction Among Fine Hematite Particles in an Exceedingly Weak Magnetic Field," *Nonferrous Metals* (in Chinese), vol. 50, no. 4, 1998, pp. 26-33.
[4] J. Fan, J. Yao, X. Zhang, K. Cen. "Modeling Particle-to-Particle Interactions in Gas-Solid Flows," *Journal of Engineering Thermophysics* (in Chinese), vol. 22, no. 5, 2001, pp: 629-632.
[5] R. Mehasni, M. Feliachi, and M. E. H. Latreche. "Effect of the Magnetic Dipole-Dipole Interaction on the Capture Efficiency in Open Gradient Magnetic Separation," *IEEE Trans. on Magnetics*, vol. 43, no. 8, 2007, pp. 3488-3493.
[6] J. Svoboda. "A Realistic Description of the Process of High-Gradient Magnetic Separation," *Minerals Engineering*, vol. 14, no. 11, 2001, pp. 1493-1503.
[7] X. Zheng, Y. Wang, D. Lu. "A Realistic Description of Influence of the Magnetic Field Strength on High Gradient Magnetic Separation," *Minerals Engineering*, vol. 79, 2015, pp. 94-101.
[8] J. H. P. Watson and P. A. Beharrell, "Extracting values from mine dumps and tailings," *Minerals Engineering*, vol. 19, 2006, pp. 1580-1587.
[9] N. Karapinar. "Magnetic Separation of Ferrihydrite from Wastewater by Magnetic Seeding and High-Gradient Magnetic Separation," *Int. J. Miner. Process*, vol. 71, 2003, pp. 45-54.
[10] E.H. Borai, E.A. El-Sofany, T.N. Morcos. "Development and Optimization of Magnetic Technologies Based Processes for Removal of Some Toxic Heavy Metals," *Adsorption*, vol. 13, 2007, pp. 95-104.
[11] L. Gao, Y. Chen. "A Study on the Rare Earth Ore Containing Scandium by High Gradient Magnetic Separation," *Journal of Rare Earths*, vol. 28, no. 4, 2010, pp. 622-626.
[12] Y. Nakai, F. Mishima, Y. Akiyama, and S. Nishijima. "Development of Magnetic Separation System for Powder Separation," *IEEE Trans. on Applied Superconductivity*, vol. 20, no. 3, 2010, pp. 941-944.
[13] N.P.H. Padmanabhan, T. Sreenivas. "Process Parametric Study for the Recovery of Very-fine Size Uranium Values on Superconducting High Gradient Magnetic Separator," *Advanced Powder Technology*, vol. 22, 2011, pp. 131-137.
[14] Y. Li, J. Wang, X. Wang, etc. "Feasibility Study of Iron Mineral Separation from Red Mud by High Gradient Superconducting Magnetic Separation," *Physica C*, vol. 471, 2011, pp. 91-96.
[15] F. Mishima, T. Terada, Y. Akiyama, and S. Nishijima. "High Gradient Superconducting Magnetic Separation for Iron Removal From the Glass Polishing Waste," *IEEE Trans. on Applied Superconductivity*, vol. 21, no. 3, 2011, pp. 2059-2062.
[16] J.H.P. Watson. "Magnetic Filtration," *J. Appl. Phys.* vol. 44, no.9, 1973, pp. 4209-4213.
[17] C. J. Clarkson and D. R. Kelland. "Theory and Experimental Verification of a Model for High Gradient Magnetic Separation," *IEEE Trans. on Magnetics*, vol. MAG-14, no.3, 1978, pp. 97-103.
[18] R. Gerber and R.R. Birss. *High Gradient Magnetic Separation*. Research Studies Press, John Wiley & Sons, Chichester, 1983.
[19] Y. G. Kim, J.B. Song, etc. "Effects of Filter Shapes on the Capture Efficiency of a Superconducting High-Gradient Magnetic Separation System," *Supercond. Sci. Technol*, vol. 26, 2013. 085002(7pp).
[20] E. Maxwell, A. Shalom, and D. R. Kelland. "Particle Size Dependence in High Gradient Magnetic Separation," *IEEE Trans. on Magnetics*, vol. 17, no. 3, 1981, pp. 1293-1301.
[21] D. Sun. "Experimental Study for Improving the Beneficiation Recovery of an Oolitic Hematite," *Mining and Metallurgical Engineering* (in Chinese), vol. 31, no. 1, 2011, pp. 43-47.
[22] Y. Hu, M. Huangfu. "High Gradient Magnetic Separation Technology Characteristics of Amphibole Hematite Ore from Yuanjiacun," *Metal Mine* (in Chinese), vol. 472, 2015, pp. 62-66.
[23] K. C. Warnke. "Finite-Element Modeling of the Separation of Magnetic Microparticles in Fluid," *IEEE Trans. on Magnetics*, vol. 39, no. 3, 2003. pp. 1771-1777.

[24] D. Bockenfeld, H. Chen, etc. "A Parametric Study of a Portable Magnetic Separator for Separation of Nanospheres from Circulatory System," *Separation Science and Technology*, vol. 45, 2010, pp:355-353.
[25] X. Zheng, Y. Wang, D. Lu. "Study on Capture Radius and Efficiency of Fine Weakly Magnetic Minerals in High Gradient Magnetic Field," *Minerals Engineering*, vol. 74, 2015, pp. 79-85.
[26] X. Zheng, Y. Wang, D. Lu. "Investigation of the Particle Capture of Elliptic Cross-Sectional Matrix for High Gradient Magnetic Separation," *Powder Technology*, vol. 297, 2016, pp. 303-310.
[27] X. Yang, and K. Chen. "Research on Occurrence State of Trace Fe and Ti in Kaolin," *China Powder Science and Technology* (in Chinese), vol. 10 Special, 2004, pp. 132-135.
[28] X. Wang, X. Lu, Y. Li. "Viscosity of Kaolin Slurry from Coal Seam and the Influencing Factors," *The Chinese Journal of Process Engineering* (in Chinese), vol. 4, no. 4, 2004, pp. 295-299.
[29] L. Zheng. "Study on Magnetic Hematite Ore," *Metal mine* (in Chinese), vol. 2, 1982, pp: 32-34.
[30] S.K. Baik, D.W. Ha, R.K. Ko, J.M. Kwon. "Magnetic Field and Gradient Analysis Around Matrix for HGMS," *Physica C*, vol. 470, 2010, pp. 1831-1836.